\documentclass{article}

\usepackage{arxiv}

\usepackage[utf8]{inputenc} 
\usepackage[T1]{fontenc}    
\usepackage{hyperref}
\hypersetup{colorlinks=true,linkcolor=red,citecolor=blue}
\usepackage{url}            
\usepackage{booktabs}       
\usepackage{amsfonts}       
\usepackage{nicefrac}       
\usepackage{microtype}      
\usepackage{graphicx}
\usepackage[square,numbers]{natbib}   
\usepackage{doi}
\usepackage{hhline}%
\usepackage{paralist}

\usepackage{titlesec} 
\titleformat{\paragraph}
{\normalfont\normalsize\bfseries}{\theparagraph}{1em}{}
\titlespacing*{\paragraph}
{0pt}{3.25ex plus 1ex minus .2ex}{1.5ex plus .2ex}

\title{
Modelling and Mining of Patient Pathways: \\
A Scoping Review
}

\author{
Caroline de Oliveira Costa Souza Rosa \\
Programa de Doutorado em Modelagem Computacional -- Laborat\'orio Nacional de Computa\c c\~ao Cient\'ifica (LNCC) \\
\texttt{carolrosa@posgrad.lncc.br} \\
\And
M\'arcia Ito \\
Programa de Mestrado em Sistemas Produtivos -- Centro Paula Souza (PPG-GTPS) \\
\texttt{marcia.ito@cpspos.sp.gov.br} \\
\And
Alex Borges Vieira \\
Departamento de Ciência da Computação -- Universidade Federal de Juiz de Fora (UFJF) \\
\texttt{alex.borges@ufjf.edu.br} \\
\And
Ant\^onio Tadeu Azevedo Gomes \\
Laborat\'orio Nacional de Computa\c c\~ao Cient\'ifica (LNCC) \\
\texttt{atagomes@lncc.br}
}

\date{}

\renewcommand{\headeright}{}
\renewcommand{\undertitle}{}

\hypersetup{
}

\begin{document}
\maketitle

\begin{abstract}
The sequence of visits and procedures performed by the patient in the health system, also known as the patient's pathway or trajectory, can reveal important information about the clinical treatment adopted and the health service provided. The rise of electronic health data availability made it possible to assess the pathways of a large number of patients. Nevertheless, some challenges also arose concerning how to synthesise these pathways and how to mine them from the data, fostering a new field of research. The objective of this review is to survey this new field of research, highlighting representation models, mining techniques, methods of analysis and examples of case studies.
\end{abstract}



\section{Introduction}

The increase in the incidence rate of chronic and acute diseases that leave sequelae in patients, such as COVID-19, has brought a challenge in how to deal with these patients. They usually have multiple chronic or long-term conditions that may include both physical and mental illnesses. For example, a patient with diabetes who contracted COVID-19 may have a motor sequel after being cured. These patients are classified as patients with multimorbidity. 
They have their daily lives greatly affected, more specifically in their quality of life, and commonly have mental illnesses, such as anxiety and depression. In addition, multimorbidity increases the fragmentation of care and affects family relationships. Thus, there is an increase in the use of the health service by patients with these conditions when compared to those who do not have multimorbidity~\cite{Poitras2018,Ito2020}.

The management of these patients in the health service becomes complex because, in addition to the complexity of dealing with various diseases, there is the problem of the intensity of interventions. Health services need to be organised to support the intensity and integration necessary for patients with multimorbidity. Furthermore, existing clinical protocols do not consider multimorbidity, and thus it is difficult for the healthcare professional to know what to do with multiple recommendations for the same patient. And for the patient and their caregivers, the decision of which recommendation to follow is also an issue. One way to develop clinical protocols and better organise the health service is to study and understand the patient pathway of patients with multimorbidity~\cite{Poitras2018}.

A \emph{patient pathway} denotes the sequence of steps followed by the patient within the health system. 
It is commonly used to evaluate the treatment provided to the patient \cite{Rinner2018,Baker2017,Gonzalez-Garcia2020}, but it is not restricted to it \cite{Dhaenens2018}. 
Due to the high amount of variables inherent to healthcare data, the type of information chosen as the focus of the patient pathway constitutes its \emph{perspective} \cite{Manktelow2022}. 
Besides treatments/interventions, patient pathway perspectives include diseases (diagnoses) \cite{DeOliveira2020b}, specialities \cite{Conca2018}, departments \cite{Arnolds2018} and organisational activities \cite{Rebuge2012}, or a combination of these
\cite{Najjar2018,Zhang2015}.  
Although obtaining and assessing such pathways is not a new topic of interest \cite{Rotondi}, the possibilities of analysing them have increased with the rise of electronic health data availability~\cite{Dahlem2015}.

\emph{Patient pathway mining} refers to the task of using computer-aided methods to examine health data, 
identify the different pathways followed by a group of patients, select relevant actions and behaviours,  and summarise them in a readable yet comprehensive way that can be used by specialists. 
It can be used, for example, to evaluate whether guidelines have been followed \cite{Rinner2018} or to propose new ones \cite {Chen2018}, to compare results of different pathways options \cite{Conca2018}, to analyse costs \cite{Dahlin2019_b} and to improve the layout of a healthcare unit \cite{Arnolds2018}. 
Although patient pathway mining has made it possible to assess the pathways of tens of thousands of patients \cite{Najjar2018,DeOliveira2020b}, 
challenges arose on how to deal with these data, especially because of their variability and complexity \cite{Rebuge2012}. 
As a result, numerous models to represent patient pathways, algorithms to mine them, and auxiliary methods to improve the results have been proposed. 

For example, taking a selection of four studies that focused on diabetic patient pathways, 
\citet{Dagliati2018} used a directed acyclic graph to represent pathways mined with a method that looks  for frequent activities and direct successions.
\citet{DeOliveira2020a} used a graph with temporal layers (time grid process model) to represent the pathways obtained with an optimisation model. 
\citet{Wang2017} took a sequence of time intervals as their patient pathway model and used frequent pattern mining to determine the typical activities within each interval. Lastly, \citet{Lismont2016} used a process mining algorithm, the Fuzzy Miner, to obtain a process map to summarise the pathways. Thus, given the multiple possibilities for exploring pathways, even for patients with the same condition, it would be interesting to review the literature on patient pathway mining to gain a better understanding of this field of research. 

In this context, \citet{Erdogan2018} and \citet{Rojas2016} reviewed studies that used process mining techniques applied to the healthcare domain. Given that a subset of process mining techniques---namely, the \emph{process discovery} ones---aim at mining a process model from event log data, they can be used to discover processes centred on the patient, i.e.\ patient pathways. 
Nevertheless, as both review articles are limited to process mining methods, they do not encompass other strategies, such as sequence mining. 
Other review articles that focused solely on process mining techniques assessed the specific fields of oncology~\cite{Kurniati2016}, disease trajectories~\cite{Kusuma2021}, and primary care~\cite{Williams2018}.

Recently, \citet{Manktelow2022} reviewed studies that derived data-driven clinical pathways. They investigated and proposed a classification of data enhancement strategies, supplemental techniques and care pathway perspectives. Nonetheless, the models and mining methods were not explicitly discussed. To the best of our knowledge, there are no reviews of studies that summarise and discuss the used methods for modelling and mining patient pathways in the literature.

The goal of this paper is to survey the literature on patient pathway modelling and mining, focusing on methods and case studies. To achieve that, we collected information about the perspectives from which patient pathways are built, the models used to represent them, how they are mined from data, the auxiliary techniques used and the case studies that were conducted with them. 

The remaining of the paper is structured as follows: in Section~\ref{sec:methods}, we describe the process of collection and selection of articles; in Section~\ref{sec:results}, we present the revised papers and address their computational and modelling aspects, as well as the medical fields and perspectives they encompass; the main findings and some perspectives for future work are discussed in Section~\ref{sec:disc}.

\section{Methods}
\label{sec:methods}

We searched scientific articles published in academic journals and written in English, using four databases: CINAHL, PubMed, Scopus, and Web of Science.
The articles were collected in March 2021 with no restriction on the publishing date. The search strategy involved two groups of words connected by the AND operator. The first one included expressions related to patient pathways, while the second included data-oriented terms: \textit{( "clinical pathway" OR “clinical treatment process” OR “patient pathway” OR “patient flow” OR “patient mobility” OR “*care trajectory” OR “trajectory of care” OR “*care pathway” OR “*care process” OR “careflow” OR “care flow” OR “medical treatment process” OR “regionalization” OR “health trajectory” ) AND ( “electronic health” OR “data driven” OR “mining” )}. The final query was adapted to each database. 

Besides this search for terms in the key fields of the articles, CINAHL and PubMed also support searches based on the subject headings they use to index pieces of work. Therefore, an extra search was performed in both databases using their subject headings. The query for PubMed using the MESH terms was: \textit{(“Critical Pathways”[Mesh] OR “Continuity of Patient Care”[Mesh] OR “Episode of Care”[Mesh] OR “Process Assessment, Health Care”[Mesh]) AND (“Electronic Health Records”[Mesh] OR “Data Science”[Mesh] OR “Big Data”[Mesh]) AND (trajector* OR pathway* OR flow)}. The CINAHL query using subject headings was: \textit{( (MH “Process Assessment (Health Care)”) OR (MH “Critical Path”) OR (MH “Continuity of Patient Care”) ) AND ( (MH “Data Science”) OR (MH “Data Mining”) OR (MH “Data Analysis”) OR (MH “Electronic Health Records”) ) AND ( trajector* OR pathway* OR flow )}.

\begin{figure}[!ht]
    \centering
    \includegraphics[width=0.99\textwidth]{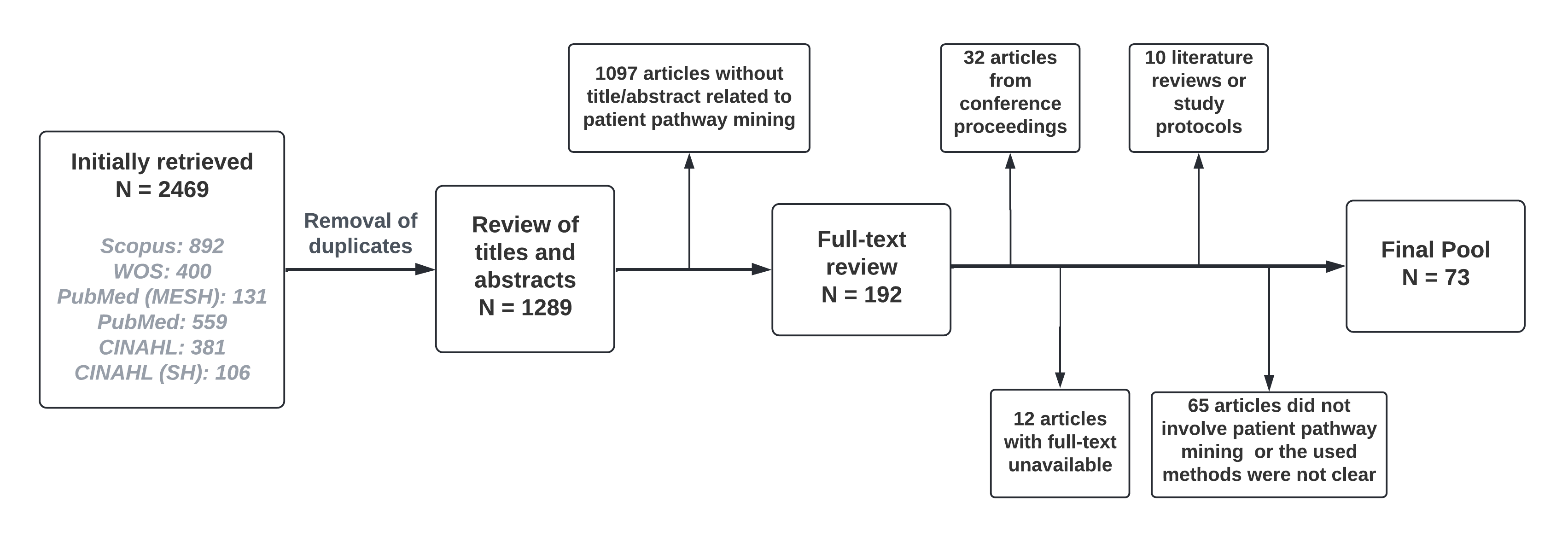}
   \caption{Process of selecting papers for this survey.}
    \label{fig:esquema2}
\end{figure}

The process of selecting articles for the review involved:
\begin{inparaenum}[(i)]
\item searching the databases; 
\item removing duplicates; 
\item reviewing titles and abstracts; and 
\item reviewing full texts.
\end{inparaenum}
The number of articles in each of these phases is shown in Figure \ref{fig:esquema2}. 
Mind that, although we chose four databases, we actually performed six searches, owing to the two different approaches used in CINAHL and PubMed. 

Initially, the searching process returned 2,469 results, which corresponded to 1,289 unique scientific articles. 
We screened their titles and abstracts to select those that mentioned or suggested the mining of patient pathways from 
healthcare data. 
After this step, we considered 192 references as relevant, so we collected and read their corresponding full texts. 
During this step, we defined the final pool of articles according to the following exclusion criteria: 
\begin{inparaenum}[(i)]
\item papers from conference proceedings or book chapters, which bypassed the filter for journal articles in the databases;
\item articles whose full text could not be accessed; 
\item secondary studies or study protocols; \item articles that did not involve patient pathway mining or whose methods were not clearly stated.
\end{inparaenum}
Finally, we selected 73 scientific articles  for this survey. 

We read the articles and drew our attention to how the patient pathways were represented and the methods used to mine them. In some of these articles, the authors mentioned and tackled the challenge of dealing with health data because of their variability; therefore, we also investigated the auxiliary methods used to deal with such variability. Moreover, we considered the number of perspectives and how time was incorporated into the pathways as other relevant features of the studies. From the medical viewpoint, we focused on medical fields whose patient pathways were explored and the perspectives upon which the pathways were built. Lastly, we also considered 
how the patient pathway mining results were employed.

\section{Results} %
\label{sec:results}

\begin{figure}[!ht]
    \centering
    \includegraphics[width=0.97\textwidth]{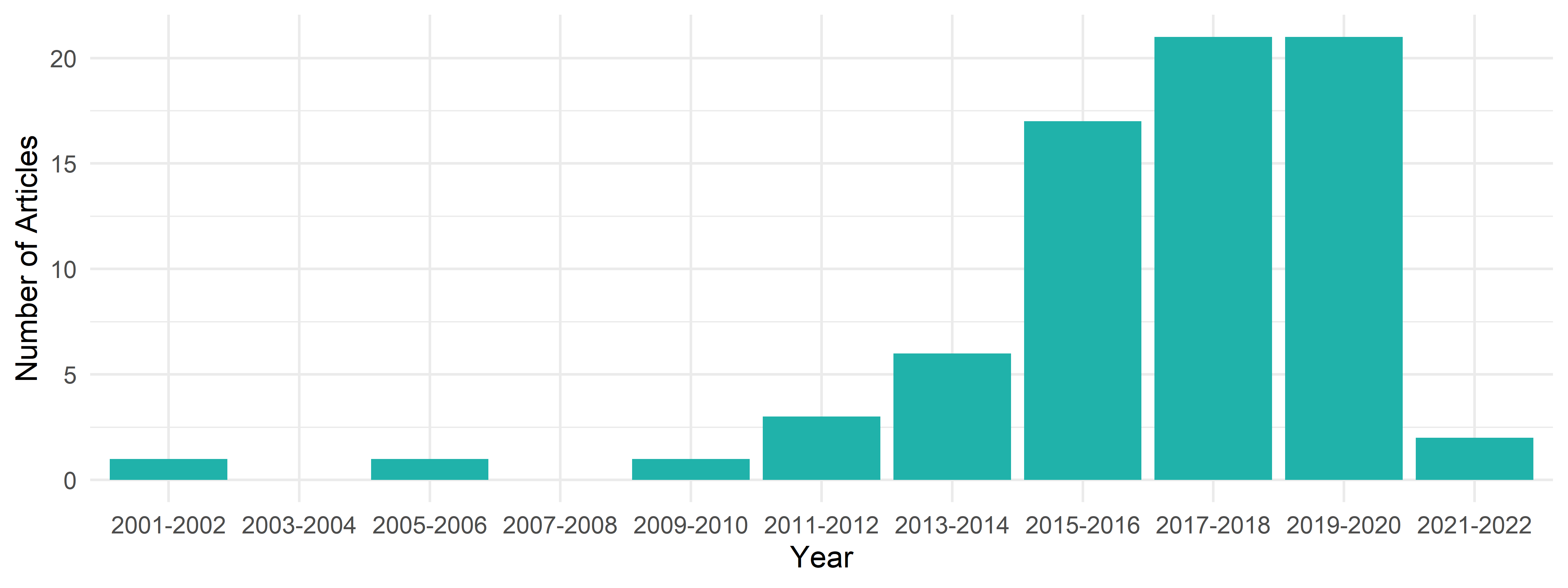}
  \caption{Frequency of articles according to publishing biennium.}
    \label{fig:year2}
\end{figure}

In this section, we present a summary of the seventy-three reviewed papers. We organised the results into two subsections. Firstly, in Section \ref{sec:modelling}, we cover the mathematical modelling and the mining methods of patient pathways. Next, in Section~\ref{sec:case}, we discuss aspects related to the medical interpretation of the pathways. Figure~\ref{fig:year2} 
shows the distribution of the selected articles in publication biennia.  

\subsection{Modelling} 
\label{sec:modelling}

When mining patient pathways, one may represent them as a list of relevant patterns or a single summarising model. Moreover, it is also important to decide the level of detail the result will bear and the type of information it will contain (e.g.\ the order of events, the time between activities and the resources involved). These choices depend heavily on the mathematical model used to represent the pathways and the algorithm chosen to discover them. In this section, we discuss the models (Section~\ref{subsec:model}) and algorithms (Section~\ref{subsec:mining}) used in the reviewed literature. Furthermore, we also present the strategies taken to deal with healthcare data variability (Section~\ref{sec:aux_dt}). Lastly, as both the model and the mining method influence how time and other perspectives of the patient pathways are represented, we also address these topics (Sections~\ref{subsec:time}~and~\ref{subsec:persp}).

\subsubsection{Pathway modelling} 
\label{subsec:model}

The authors of the reviewed papers have modelled patient pathways in several ways. The choice of the model depends on the type of information contained in the trajectory, the goal of the study and the sought relations between activities. 

Firstly, a straightforward manner to represent pathways is using sequences of events \cite{Antonelli2012, Aspland2021, Defossez2014, Egho2014, Fei2013, Huang2012, Hur2020, LeMeur2015, Perer2015}. 
With this approach, it is possible to analyse features of the different pathways (variants) and of the patients who followed them.  Even when the authors are interested in building a more sophisticated patient pathway model, such as in process mining studies, it is common to complement the process discovery task with the inspection of the most frequent sequences (traces)~\cite{Kempa-Liehr2020,Gonzalez-Garcia2020,Kim2013,Durojaiye2018,Andrews2020,Kurniati2019a}.
However, if the variability of pathways is high,\footnote{i.e.\ the number of different pathways is high when compared to the number of patients} most of them will be followed by only a few or even by a single patient. Therefore, sequences might also depict frequent sub-patterns of the pathways, which can be obtained with sequence mining algorithms, for instance.

Regardless of using the original pathways or the most frequent patterns, there will be a list of sequences to be analysed. Such lists suit well studies that intend to use patient pathways as input for optimisation algorithms \cite{Arnolds2018} or predictive tools \cite{Hur2020,Kempa-Liehr2020}.
Nevertheless, when there is interest in understanding the overall behaviour of patients, it might be impractical for specialists to analyse every possible pathway. Alternatively, a model that summarises multiple individual trajectories, i.e.\ a model capable of replicating the observed trajectories or most of them, can represent the patient pathways concisely.  
Examples of such generic models are graphs, Markov chains and process algebras.

A graph is a mathematical structure that represents a set of objects (nodes) and the relationships between them (edges). In the patient pathway domain, for instance, the nodes could denote the encounters, while the edges would indicate which ones are directly followed by the others. Some process mining algorithms, such as the Fuzzy Miner~\cite{Gunther2007}, generate a graph as their process model. Frequently, they include virtual nodes to indicate the beginning and the end of the pathways
~\cite{Lin2001,Sato2020,Najjar2018}. Graphs can represent the whole dataset \cite{Basole2015}, or when there are many pathway variants, they might display frequent or relevant behaviour~\cite{Lin2001,Zhang2015,Prodel2018}. Moreover, the edges can bear information about the time or the patient flow between events~\cite{Gonzalez-Garcia2020,Arias2020}. 

Traditionally, the nodes of a graph have unique labels. Thus, when an activity (node) repeatedly appears in the pathways, the graph will have cycles that might hinder the interpretation of the model, especially the notion of which previous nodes were visited before the current one.
As an alternative, \citet{DeOliveira2020a} have proposed a graph with multiple layers, each one corresponding to a position in a trace. As a result, an activity can appear multiple times in the model, provided that the occurrences are in different layers. 
\citet{Dagliati2017} have used a directed acyclic graph (directed tree graph) to model the patient pathways and, therefore, repeated labels are allowed to prevent the emergence of cycles. The model proposed by~\citet{Duma2020}, named Hybrid Activity Tree, starts by adding all observed pathways into a directed tree graph, but afterwards, the algorithm converts infrequent branches into a directed cyclic graph with no repeated labels.  

Some authors have also modelled patient pathways as a set of states, with a particular transition probability between them. Markov Models~\cite{Baker2017,Garg2009,Villamil2017}, Hidden Markov Models~\cite{Zhang2015a} and Probabilistic Deterministic Finite State Automata~\cite{Arnolds2018} are examples of such approach.

When the authors expect the patient pathways to be the result of a well structured underlying process, models that support process notation (e.g.\ AND/OR/XOR joins and splits) are suitable.  
The reviewed papers included BPMN~\cite{Kurniati2019a,Lu2016,Stefanini2020a,Tamburis2020}, HeuristicNet~\cite{Caron2014,Durojaiye2018}, Inductive Visual Model~\cite{Andrews2020,Kempa-Liehr2020,Marazza2020}, Petri Nets~\cite{Durojaiye2018,Marazza2020,Rebuge2012,Stefanini2020a,Tamburis2020}, Fork/Join Networks~\cite{Senderovich2016}, and declarative process models~\cite{Mertens2018,Mertens2018}. 

Another way of modelling a patient pathway is using a sequence of time intervals, each of which has a set of frequent activities. The authors who used this strategy were mainly interested in discovering actually-adopted clinical pathways~\cite{Cho2020,Huang2013,Wang2017}.
The task may involve not only determining the set of actions, but also the time intervals.

\subsubsection{Pathway mining techniques} 
\label{subsec:mining}

\begin{figure}[!ht]
    \centering
    \includegraphics[width=\textwidth]{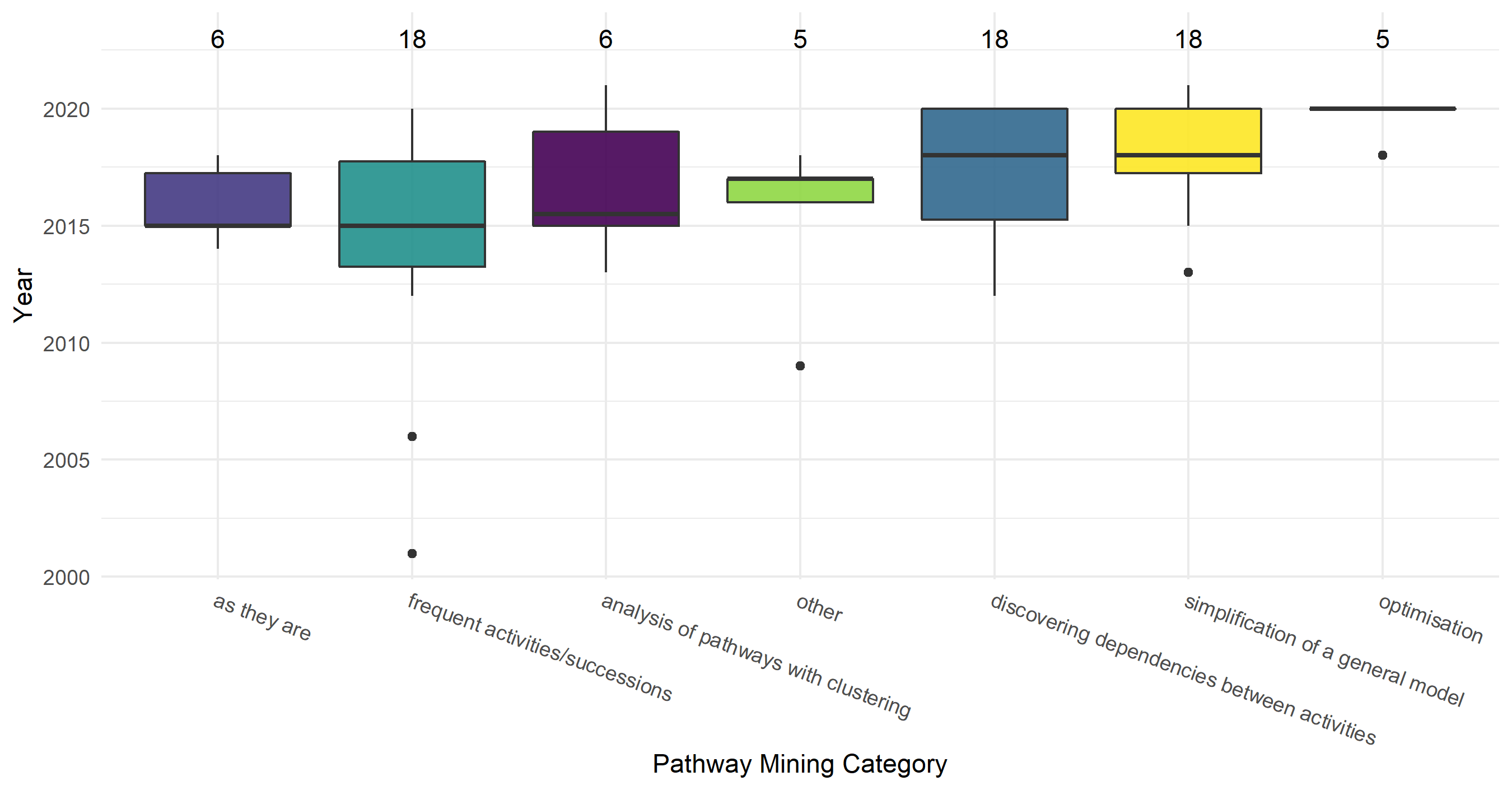}
    \caption{Publication year distribution according to Pathway Mining strategy. The number of articles in each category is present in the upper part of the chart.}
    \label{fig:dt_year}
\end{figure}

\begin{figure}[!ht]
    \centering
    \includegraphics[width=\textwidth]{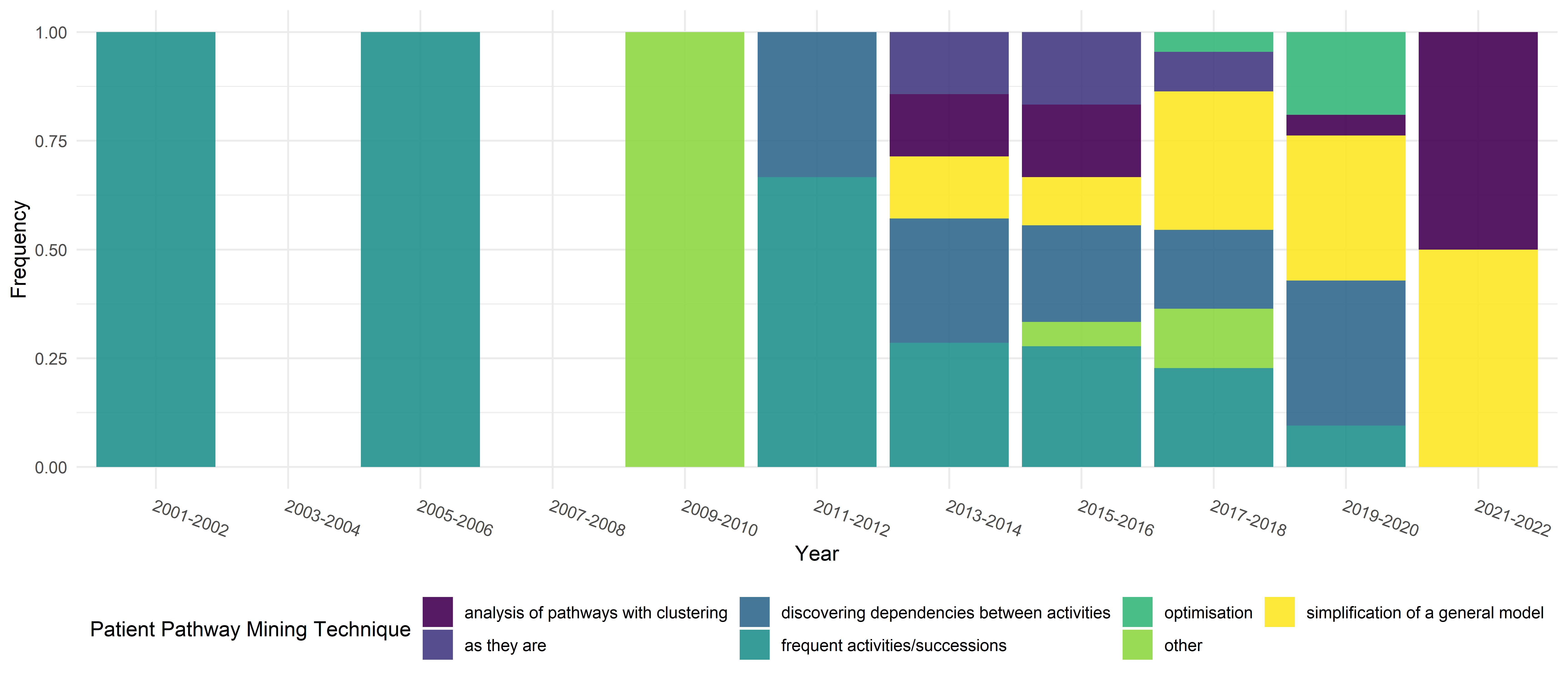}
    \caption{Distribution according to Pathway Mining strategy according to publishing biennium.}
    \label{fig:dt_year_2}
\end{figure}

In this section, we present the methods used for pathway mining.  We have grouped the reviewed papers into seven categories according to their mining strategy and will discuss them in the following subsections. Figures \ref{fig:dt_year} and \ref{fig:dt_year_2} show the distribution of publication years for each category. 

\paragraph{Algorithms based on frequent activities/successions}
\label{par:dt_freq}

The largest group of papers encompasses different methods, either proposed by the authors or previously published in the literature. They have in common the fact that events and transitions are only selected for the final model if they are frequent enough. 

Sequence mining algorithms, also called frequent or sequential pattern mining algorithms, aim at finding frequent subsequences in the data according to a chosen minimal support threshold.
They are mainly based on the strategy of the Apriori algorithm \cite{Apriori}, proposed to discover frequent sets of items and later adapted for sequential patterns \cite{AprioriAll}. 
With this approach, instead of checking the prevalence of every possible sequence, only those sequences whose subsequences are frequent are considered. For example, given three activities \textit{A}, \textit{B} and \textit{C}, the pattern \textit{ABC} can only be frequent if both subsequences \textit{AB} and \textit{BC} are frequent.
There are multiple implementations of sequence mining algorithms in the literature. The authors of the reviewed papers have used the BIDE algorithm \cite{Antonelli2012} and the CloFAST algorithm \cite{Hur2020}. Moreover, \citet{Perer2015} modified the SPAM algorithm to include a pattern duration threshold to select frequent patterns whose length of time is below a specified limit. 

In some of the reviewed studies, the authors have opted to implement a new sequence mining algorithm to deal with patient pathways. For example, \citet{Egho2014} considered that each encounter in a patient trajectory may contain multiple records of different types (e.g.\ set of procedures, diagnoses, medical speciality) and proposed a new sequence mining algorithm to discover multidimensional patterns and, in posterior work, a similarity metric to compare such sequences \cite{Egho2015}. While this piece of work highlighted how patient pathways might involve multiple perspectives, \citet{Huang2012} drew attention to incorporating time information into the mined patterns. The authors proposed a new sequence mining algorithm, whose last step converts the frequent patterns into chronicles. In other words, besides mining frequent subsequences, the authors also mine key time intervals between directly or not directly-followed activities.

\citet{Huang2013} and \citet{Wang2017} included temporal information in the models differently. They proposed a method to mine patient pathways expressed as a sequence of time intervals, each one with a set of frequent actions. They used optimisation to define the best intervals and then applied algorithms based on the Apriori \cite{Huang2013,Wang2017} and the FP-growth \cite{Wang2017} algorithms to discover frequent patterns. These interval-based patient pathways resemble how some clinical pathways are defined, and indeed, both studies had the objective of identifying actually-adopted clinical pathways.

A disadvantage of sequence mining algorithms is that sequences have limited expressive power, as they can only represent direct successions of activities. Alternatively, \citet{Lin2001} and \citet{Yang2006} proposed a modification of sequential pattern mining algorithms to discover frequent subgraphs. The use of graphs allows the depiction of concurrent activities in the pathway model. \citet{Lin2001} also included an option to define time windows and mine the pathway patterns in each one.   

The algorithm proposed by \citet{Chen2018} includes both interval-based pathways and graphs. The pathway discovery method starts with defining the time intervals according to specialists. Next, an algorithm identifies the activities undertaken in each interval and builds, for each patient, transition matrices between pairs of intervals. After that, sequences of transition matrices represent the patient pathways. The authors proposed a similarity metric to compare those sequences and group similar patients. For each cluster, they identified frequent transitions (`dense core') and included them in a graph to represent the frequent behaviour of the group. 

This approach of grouping patients and identifying frequent activities and transitions between them have been used by other authors as well. 
\citet{Najjar2018} proposed the use of two clustering steps to group patients. In the first one, they used Hidden Markov Models to cluster the pathways, and in the second, they used hierarchical clustering to group the clusters from the first phase. Then, they built a graph to depict the frequent behaviour of each final cluster. On the other hand, \citet{Zhang2015} and \citet{ZhangPadman2015} started by grouping patients with hierarchical clustering, using a similarity measure based on the longest common subsequence. They built a Markov Model using transitions of visits as states (`super pair') for each cluster. The states and transitions whose frequency was higher than a chosen threshold were used to build a graph, i.e.\ the final pathway model. The authors have also taken a similar approach but using Hidden Markov Models to predict future events \cite{Zhang2015a}. 

Although graphs can represent concurrent activities, they traditionally do not allow nodes to have repeated labels, and consequently, if the pathways have many repeated activities, the graphs end up having many loops and cycles and become hard to read. Alternatively, the method proposed by \citet{Dagliati2017} begins with the selection of frequent first activities and builds a tree adding branches whose frequency is higher than a predetermined threshold. The authors added temporal information (e.g.\ median time) to the nodes and transitions as well. They compared the values of selected biomarkers of patients who followed different pathways. In a following paper \cite{Dagliati2018}, the authors proposed an extension of the method to identify concurrent activities. 

The studies we introduce in this section encompass a broad range of publication years, as shown in Figure \ref{fig:dt_year}. The two oldest papers belong to it, and it has the second lower median year. Nevertheless, it also contains quite recent studies.   

\paragraph{Algorithms based on discovering dependencies between activities}

The development of process mining, a research field that involves algorithms and methods to discover and assess processes using event log data, has innovated the pathway mining area. One of the main advantages of such algorithms is the ability to discover and represent more types of relations between activities, e.g.\ long-term dependencies, mutually-exclusive activities, and parallel activities. 

The Heuristics Miner~\cite{heuristicsMiner}, one of the first process mining algorithms, builds a process model by identifying dependency relations between activities, taking into account a measure of their frequency. Several authors have used this algorithm for discovering patient pathways. In 2012, for instance, \citet{Rebuge2012} proposed a methodology to apply process mining in healthcare. They used sequence clustering to group patients according to their pathways and evaluated which clusters correspond to frequent/infrequent behaviour. Then, they obtained a process model with the Heuristics Miner Algorithm for each selected cluster. \citet{Caron2014} also proposed a selection of process mining techniques to assess treatment pathways. They used the Heuristics Miner algorithm in the process discovery phase. \citet{Kurniati2019a} used the algorithm as well while assessing the quality of the MIMIC-III database for process mining studies. \citet{Leonardi2018} also applied it to discover patient pathways after a pre-processing step proposed by the authors to organise and group events.
\citet{Partington2015} aimed at comparing the clinical pathways adopted by four hospitals, and they used the Heuristics Miner algorithm to discover a model for each unit.
\citet{Yoo2015} reported that they used several process mining algorithms, including the Heuristics Miner, to discover actually-followed clinical pathways. The results were compared to the theoretical clinical pathway using matching rates proposed by the authors. 
\citet{Yoo2016} used the Heuristic Miner and other process mining tools to compare the patient pathways before and after the construction of a new building in a hospital. \citet{Durojaiye2018} used the Flexible Heuristics Miner to map the pathways of pediatric trauma patients. 

More recently, \citet{InductiveMiner} proposed the Inductive Miner algorithm. It decomposes the observed sequences into subsets, and the subsets are further partitioned to reveal the underlying structure of the process. The algorithm organises the results in a process tree that can be converted to other process modelling notations. It was the algorithm used by \citet{Andrews2020} to discover the pathways of patients who suffered a road accident. \citet{Stefanini2020a} chose the algorithm to discover lung cancer pathways. They used the resulting model to support an assessment of the resources needed to treat those patients. While assessing emergency department processes, \citet{Stefanini2018} reported they mined the patient pathways using both the Inductive Miner and the Fuzzy Miner algorithms (Section
~\ref{par:dt_simp}). 

Patient pathway models not only provide a better understanding of a specific group of patients but allow comparing it with another cohort. 
\citet{Marazza2020} proposed a method to compare pathways of two sub-population of cancer patients. The idea lies in obtaining a process model for each group, converting the process models to direct graphs, and comparing them with Graph Edit Distance or with a vector of graph attributes. They used the Inductive Miner to obtain pathways represented as Petri nets, and also an extension of the algorithm---the Inductive Miner Infrequent---to mine pathways represented with the Inductive Visual Model. This latter algorithm was also the one \citet{Tamburis2020} chose to discover cataract treatment pathways. They used the process model to create and run a simulation model. Similarly, \citet{Kempa-Liehr2020} applied the Inductive Miner Infrequent to discover the pathways of patients with appendicitis. They used the most frequent pathways as attributes for predicting the postoperative length of stay. 

Most authors in this category of papers used established process mining algorithms to mine patient pathways. Nevertheless, some authors have proposed new ones focusing on the characteristics of healthcare data. For example, \citet{Lu2016} proposed a new algorithm able to discover Decomposable Cyclic Dependencies, that happen, for instance, when two activities seem to be concurrent, but an attribute value determines which one should happen first. They conducted a case study considering an organisational healthcare process. \citet{Mertens2018} presented another algorithm to mine declarative process models. It focuses on discovering rules that externalise tacit knowledge rather than producing a structured process model.  It mines the rules in two steps. The first one is based on the Apriori algorithm, and the second uses a genetic approach. The authors conducted a case study considering arm-related fractures. In further work \cite{Mertens2020}, they proposed a methodology for applying declarative process mining in healthcare based on their algorithm.

Figure~\ref{fig:dt_year} shows that mining dependencies between activities to discover patient pathways is a common approach, and tends to occur in more recent papers when compared to the methods based on frequent activities and transitions. 

\paragraph{Algorithms based on the simplification of a general model}
\label{par:dt_simp}

Process mining traditionally aims to discover process models from event log data assuming there is a well-structured underlying process, however, this assumption does not always hold. The pathways followed by patients depend on personal characteristics, medical knowledge and resource availability. Therefore, the discovered models represent so many behaviours that it becomes hard to read and interpret. Some authors have proposed alternative algorithms to avoid these so-called 'spaghetti' models \cite{Gunther2007}. The Fuzzy Miner Algorithm \cite{Gunther2007} initially translates activities and transitions between them into nodes and edges of a directed graph. Next, edges are filtered, and nodes are kept, clustered or removed. The selection of nodes and edges is not based exclusively on their frequency, but other measures, such as the relative importance of an edge for a node, and the position of the node in the network (e.g.\ a node that represents a fork in the model might be important to consider) can be used. Three research papers used the Fuzzy Miner \cite{Lismont2016,Kim2013,Partington2015}, and \citet{Xu2017} used a method based on it \cite{Xu2016}. The process discovery algorithm implemented in the software Disco\textsuperscript{\textcopyright} is based on the Fuzzy Miner, and several authors have used it to mine patient pathways
~\cite{Benevento2019,Dahlin2019_b,Erdogan2018b,Rismanchian2017,Sato2020,Rinner2018,Stefanini2018,Sawhney2021a}. Similarly, Celonis\textsuperscript{\textcopyright}, another commercial process mining tool is also based on the Fuzzy Miner \cite{Lira2019} and has been used by \citet{Arias2020}.
Furthermore, bupaR\textsuperscript{\textcopyright}~\cite{bupar},a process mining package for R, has a function to plot a graph (process map) with the possibility of applying multiple filters to simplify it. This package has been used by some authors as well~\cite{Prokofyeva2020a,Gonzalez-Garcia2020}. 

Besides the approaches based on the Fuzzy Miner algorithm, \citet{Duma2020} proposed a new pathway model (Hybrid Activity Tree) and the corresponding discovery algorithm. It starts by building a tree whose nodes represent the activities in the pathways. As no filter is applied initially, all observed behaviour is included in the tree, and it does not allow any unobserved behaviour. Afterwards, the algorithm converts infrequent branches into direct graphs to increase the generalisation of the model. Therefore, the initial model is simplified into a more general one. In addition, \citet{Arnolds2018} used a tree-like structure to represent general behaviour (Probabilistic Prefix Tree Acceptor). In this model, the nodes represent states, and the branches represent activities. The algorithm adds every observed behaviour to the model and then merges the states according to the similarity of the frequency distribution of their outgoing branches. Similarly, the PALIA algorithm
~\cite{Fernandez2008} used by \citet{Conca2018} starts building a Parallel Acceptor Tree with the activities while identifying their parallelisms, and the nodes and branches can be fused or deleted to simplify the model afterwards.   

This group of papers that simplify a general model to obtain the final one corresponds to the second largest group in Figure~\ref{fig:dt_year}. The articles tend to be more recent than the ones in the group of dependencies. 

\paragraph{Algorithms based on optimisation}

This fourth group of studies is similar to the latter one in the sense that it also chooses the most important components for the final model. However, instead of simplifying an initial model, an optimisation problem is defined to obtain the model based on some desired criteria.

In 2018, \citet{Prodel2018} proposed a novel pathway mining algorithm based on maximising the \textit{replayability} of the discovered model. The replayability metric measures the ability of the model to reproduce the observed behaviour, and one can define it in several ways. Particularly, the authors used eight different metrics. The output of the optimisation problem is a directed graph, and the tabu-search algorithm used to solve it decides which activities will be represented/grouped in a node and which edges will be included in the model. While the objective function aims at maximising the fitness of the model, a threshold is established to limit the total number of nodes and edges to avoid an overly complex model. A later study used the method in another case study \cite{DeOliveira2020b}.

Based on this approach, but aiming at providing a more representative model, the same research group extended the method to mine graphs with layers and edges characterised by time intervals~\cite{DeOliveira2020a}. 
Such graphs allow recurring events to appear in different layers of the model. Moreover, they support multiple transitions between the same pair of nodes to represent a set of distinct time intervals.
The replayability metric considers the percentage of replayed events, the proportion of missing edges or unsuitable time intervals, and whether there were skipped events. There are constraints on the number of nodes, edges and layers. In a further publication, the authors proposed an event clustering preprocessing step and used the method to mine the pathways~\cite{DeOliveira2020c}.

\citet{Cho2020} developed another pathway mining algorithm based on optimisation. Their objective was to decide which orders, i.e.\ direct succession between two activities, should be included in the model to maximise its fitness and precision, according to two proposed matching rates. The algorithm obtains the optimal model checking iteratively if the inclusion of an order increases the average matching rate or not.   

Although this is the category with the smaller number of papers, it is also the group with the highest median (i.e.\ most recent). 
 
\paragraph{Analysis of pathways with clustering}
\label{par:dt_clust}

Some of the reviewed papers assessed patient pathways based on clustering them and analysing the characteristics of each group, without explicitly mining relevant patterns or building models to represent them.
Although this survey focuses on studies involving patient pathway modelling, we considered such an approach relevant to be discussed and thus, we present the papers in this section. 

Firstly, \citet{Huang2015} used a method based on Latent Dirichlet Allocation to group patients into patterns of similar treatment pathways. 
It allows, for instance, the evaluation of the frequency of an activity throughout the treatment days for each pattern. \citet{Huang2015a} applied the method to detect local anomalies in clinical pathways. In posterior work, they have enriched the method to consider initial comorbid illnesses as features for the clustering step \cite{Huang2016}. 

Several authors opted for k-medoids clustering. \citet{Zaballa2020} used it and chose the Edit Distance to measure the similarity between the pathways. They considered the medoid of each cluster as its representative behaviour and specialists assessed the results. On the other hand, the objective of \citet{Aspland2021} was to propose a new similarity metric, which penalises differences between the sequences according to the case study and the activities groups. \citet{Fei2013} used two clustering steps based on k-medoids clustering. The first one is based on pathway features, while the second considers the sequence of actions.  

This group of papers includes recent references, but its median publication year is close to the median of the group of articles based on frequent activities/sequences. Mind that, though, that clustering has also been used as an auxiliary method for other discovery categories, as will be discussed in Section~\ref{sec:aux_dt}.

\paragraph{Extraction of patient pathways as they are}

Some authors used neither a pathway mining algorithm nor a clustering method. This happened mainly when the case study and the pathway construction led to few variants or when there was a preprocessing step that simplified the pathways significantly.

Examples of simple pathways are the ones considered by \citet{Zhang2018}. They consisted of changes of medicine for treating specific conditions, and they presented the results with charts. Furthermore, the patient pathways modelled by \citet{LeMeur2015} had exactly three events, each one summarising the prenatal care of the corresponding trimester of pregnant patients. 

When it comes to preprocessing for simplifying pathways, \citet{Huang2018} proposed a method to cluster pathway events into treatment topics.  Although they provided a visual representation of only a few selected pathways in the study, they mentioned the intention to process the overall topic trajectories in future work. Another example is the work of \citet{Defossez2014}, in which they converted groups of events of breast cancer trajectories into states according to a set of rules.  

In their work, \citet{Bettencourt-Silva2015} included several steps of filtering and clustering events according to the case study. They developed a system for specialists to assess the results. It included, for example, a chart showing the pathway of clinical activities versus the evolution of a biomarker measured for a patient.
\citet{Basole2015} also focused on providing a visual representation of patient pathways. They used a graph and strategically arranged the nodes to facilitate its interpretation.

The seven articles in this group correspond to the oldest median year in Figure~\ref{fig:dt_year}.

\paragraph{Others}

The five following papers did not suit well in the previously presented categories of patient pathway mining techniques but did not constitute another group by themselves. Therefore, we introduce them in this section.

Two studies share the fact that they start with a predefined pathway model. One of them updates the model and enriches it while reviewing the event log~\cite{Baker2017}, while the other calculates the model parameters with the observed data~\cite{Garg2009}. We present more details of each one of them below. 
\citet{Baker2017} started with a Markov model proposed by specialists to depict the patient pathways.  They obtained the transition probabilities with data from the observed pathways and updated the states of the model if an unforeseen event appeared in the data.  Moreover, they identified how many patients had each state as part of their most severe sub-pathway. \citet{Garg2009} began with a Markov Model as well and estimated the transition probabilities with the available data. They identified, among others, the most likely, the most expensive, and the shortest pathways.

\citet{Villamil2017} chose a different approach. They used an originally clustering algorithm as their pathway mining method, setting the number of clusters to one. The method starts by generating a predefined number of Markov models. In each iteration, it assigns every patient pathway to the model most likely to reproduce it and updates the parameters of the models. It repeats these steps until the parameters of the models converge. In the end, it obtains a Markov model to represent each group. 

Focusing on diagnosis pathways, \citet{Khan2018} used complex networks to represent the evolution of comorbidities in diabetic patients. They prepared a disease network for non-diabetic patients and another for diabetic patients. They compared the prevalence of diseases (nodes) in each network and identified those that were more prevalent for diabetic patients.

Lastly, to evaluate the conformance of cancer patient pathways, \citet{Senderovich2016} used an approach based on interval algebra and Markovian probabilities to discover the structure of the model and characterise its dynamics to build a Fork-Join network that describes the process. 

\subsubsection{Strategies for dealing with data complexity} 
\label{sec:aux_dt}

Many authors have highlighted the difficulties that arise from the large variability of healthcare data when mining patient pathways. Therefore, they have used several auxiliary methods, which we present in this section. 

Firstly, the most frequently reported auxiliary method was filtering events. The majority of the authors had a list of relevant (or irrelevant) events for the case study and automatically removed them from the data. Alternatively, some authors filtered events based on their frequency~\cite{Wang2017,DeOliveira2020c}. \citet{Zaballa2020} proposed a method for selecting relevant events when the dataset is incomplete. They filtered events according to the corresponding diagnosis but, when this information was not available, they made the decision based on the corresponding speciality, comparing its frequency in the relevant and irrelevant groups of events. They highlighted the importance of this selection because one patient might follow multiple pathways (referring to different diseases) simultaneously, and they may overlap in the dataset.

Filtering events might prevent authors from finding interesting patterns. Indeed, some authors purposely chose not to filter them to investigate how comorbidities or other illnesses could influence the studied pathway~\cite{Najjar2018,Prodel2018,DeOliveira2020a}. To cope with that, \citet{Najjar2018} clustered events and trajectories, while \citet{Prodel2018} and \citet{DeOliveira2020a} used an optimisation model to select nodes and edges to mine the pathways.

Another usual auxiliary method is clustering events. It can result, for example, from the specialists' guidance~\cite{Chen2018,Dagliati2018}, or the identification of low-level activities~\cite{Erdogan2018b,Stefanini2020a}. Moreover, when the activities have a label with a hierarchical code, such as the International Classification of Diseases, some authors opted to use a less specific category so that the specific ones formed a group~\cite{Marazza2020,Villamil2017}.

Researchers have also used numerous automatic methods to cluster events, including autoencoding~\cite{DeOliveira2020c}, topic models~\cite{Chiudinelli2020, Xu2017, Huang2018}, DBScan~\cite{Antonelli2012} and k-prototypes~\cite{Najjar2018} clustering algorithms, and itemset mining~\cite{Perer2015}. \citet{Leonardi2018} proposed an algorithm that maps activities to an ontology and identifies macro-actions, i.e.\ sets of activities that happen close to each other and refer to the same objective. 

Besides event clustering methods, patient clustering can support pathway discovery likewise. Authors have adopted decision trees~\cite{Duma2020}, k-medoids~\cite{Aspland2021,Zaballa2020}, hierarchical~\cite{Zhang2015a,ZhangPadman2015,Zhang2015} and Active Trace~\cite{Lismont2016} clustering algorithms.  \citet{Rebuge2012} used a clustering method based o Markov models. The methods proposed by \citet{Najjar2018}, \citet{Fei2013} and \citet{Chen2018} also involve clustering and were detailed in Sections~\ref{par:dt_freq}~and~\ref{par:dt_clust}.

Another auxiliary method is filtering pathways, either based on their frequency~\cite{Rismanchian2017,Tamburis2020} or their duration (length of stay)~\cite{Wang2017}. Additionally, the authors who adopted the Inductive Miner had the option of manually editing the discovered process model. \citet{Andrews2020} and \citet{Stefanini2020a} used this aproach. 

Some studies that did not mention the use of an auxiliary method involved shorter pathways, such as acute illness pathways \cite{Cho2020,Kempa-Liehr2020,Yoo2015,Gonzalez-Garcia2020,Sato2020,Sawhney2021a}, emergency department pathways~\cite{Basole2015,Benevento2019,Partington2015}, and pathways of medicine alternation~\cite{Zhang2018}. Other studies comprised department/resource~\cite{Arnolds2018,Durojaiye2018,Senderovich2016} or organisational pathways~\cite{Arias2020,Kim2013,Lu2016}. 
Lastly, for some studies, the pathway mining method involved some type of filtering or clustering, e.g.\ sequence mining~\cite{Huang2012,Huang2013,Hur2020} or optimisation~\cite{Prodel2018,DeOliveira2020a}.

\subsubsection{Time} 
\label{subsec:time}

The analysis of patient pathways naturally involves a temporal perspective, as they consider the order of the events the patients followed. However, one can further explore it by assessing the time between events. We identified multiple approaches to enrich the time analysis and summarise them in this section.  

First of all, many authors have included temporal information, such as mean, median, maximum and minimum time, in the edges (transitions) and nodes (activities) of the process model~\cite{Andrews2020,Chiudinelli2020,Erdogan2018b,Dagliati2018,Prodel2018}. It enables statistical and performance analyses, including the identification of bottlenecks~\cite{Andrews2020,Benevento2019} and the evaluation of conformity to temporal guidelines
~\cite{Gonzalez-Garcia2020,Sato2020}.

Moreover, the temporal information might be part of the pathway models or their mining algorithms. For instance, \citet{Leonardi2018} considered the timestamp of the events to decide which ones corresponded to the same macro-action. Additionally, \citet{Defossez2014} converted the observed events into states and considered both the sequence of the states and a sequence whose states are repeated proportionally to their duration. The timestamps can also support clustering of close events~\cite{Antonelli2012,Baker2017} or be at the basis of the pathway definition, e.g.\ in \cite{Rinner2018}, the events differ from each other by the number of the visit and the occurring trimester, and in \cite{LeMeur2015}, the authors built the pathways over prenatal trimesters. 

\citet{Lin2001} allowed the partition of the pathways in different time windows.  
Similarly, studies considering interval-based pathways can easily store temporal information because the intervals are time-based. Moreover, they benefit from the fact that an activity can appear in different moments of the same patient pathway. Although \citet{Cho2020} and \citet{Bettencourt-Silva2015} did not use intervals, they achieved a similar result because they defined the pathway events as a combination of the activity and the occurring day (relative to a baseline event, such as diagnosis or surgery), besides other information. Thus, the events keep temporal information and, when comparing pathways, the occurring time could differentiate them. Similarly, the events defined by \citet{ZhangPadman2015} (also in \cite{Zhang2015a}) used temporal categories (e.g.\ less than three months, three to six months, more than six months) to define the super pairs (two visits in sequence) used as states of their Markov model. 

The grid process model proposed by \citet{DeOliveira2020a} (also in \cite{DeOliveira2020c}) is a graph whose nodes contain the activity and its position in the pathway. Furthermore, the mining algorithm discovers relevant temporal intervals for the edges so that the same edge, i.e.\ sequence of two activities, can represent different behaviours according to the time elapsed between them. \citet{Huang2012} also worked with temporal constraints, but in this case, the base model is a sequence, and the time intervals can include more than two activities.


\subsubsection{Perspectives} 
\label{subsec:persp}

Most authors built patient pathways considering a single perspective, e.g.\ diagnosis pathways or speciality pathways. However, some opted for multiple-perspective pathways.  Authors who used hospitalisation data could have, for instance, diagnosis, intervention and medical unit data \cite{Egho2015}. The goal of \citet{Egho2014} was precisely to provide a method for mining sequences of  multidimensional item sets. They observe the hierarchical level of each category of data to discover frequent patterns.  

Other approaches include clustering multi-perspective events \cite{Najjar2018} and creating labels with concatenated information \cite{Zhang2015,Zhang2015a,ZhangPadman2015}, or autoencoding \cite{DeOliveira2020c}. Alternatively, in some studies, the patient pathways encompassed a single perspective, but a second one played an auxiliary role. For example, \citet{Fei2013} studied speciality pathways but, when comparing two sequences, they used the departments where two visits took place to measure the distance between them. Furthermore, \citet{Zaballa2020} used speciality information to estimate whether an event with a missing diagnosis was likely to be related to the case study.  

A couple of authors worked with qualitative and quantitative events. \citet{Bettencourt-Silva2015} built intervention pathways and compared them with a biomarker time series. They highlighted that biomarker trends can be used to assess the quality of the recorded intervention data. Similarly, \citet{Conca2018} labelled patients according to the evolution of a blood laboratory test result and assessed whether these labels correlated with the groups of patients clustered according to the speciality pathway. On the other hand, \citet{Dagliati2017} used biomarkers measured at strategic points to compare different groups of patient pathways. Furthermore, when \citet{Huang2018} grouped events into topics, they considered both the activity and its intensity, e.g.\ the number of times an intervention was done or a medicine dosage.


\subsection{Case study} 
\label{sec:case}
In this section, we address the papers according to their case study. We considered which medical fields the patients belonged to, the kind of information contained in the pathways (their perspective), and the objective of the patient pathway mining study. 

\subsubsection{Medical fields}

\begin{figure}[!ht]
    \centering
    \includegraphics[width=0.95\textwidth]{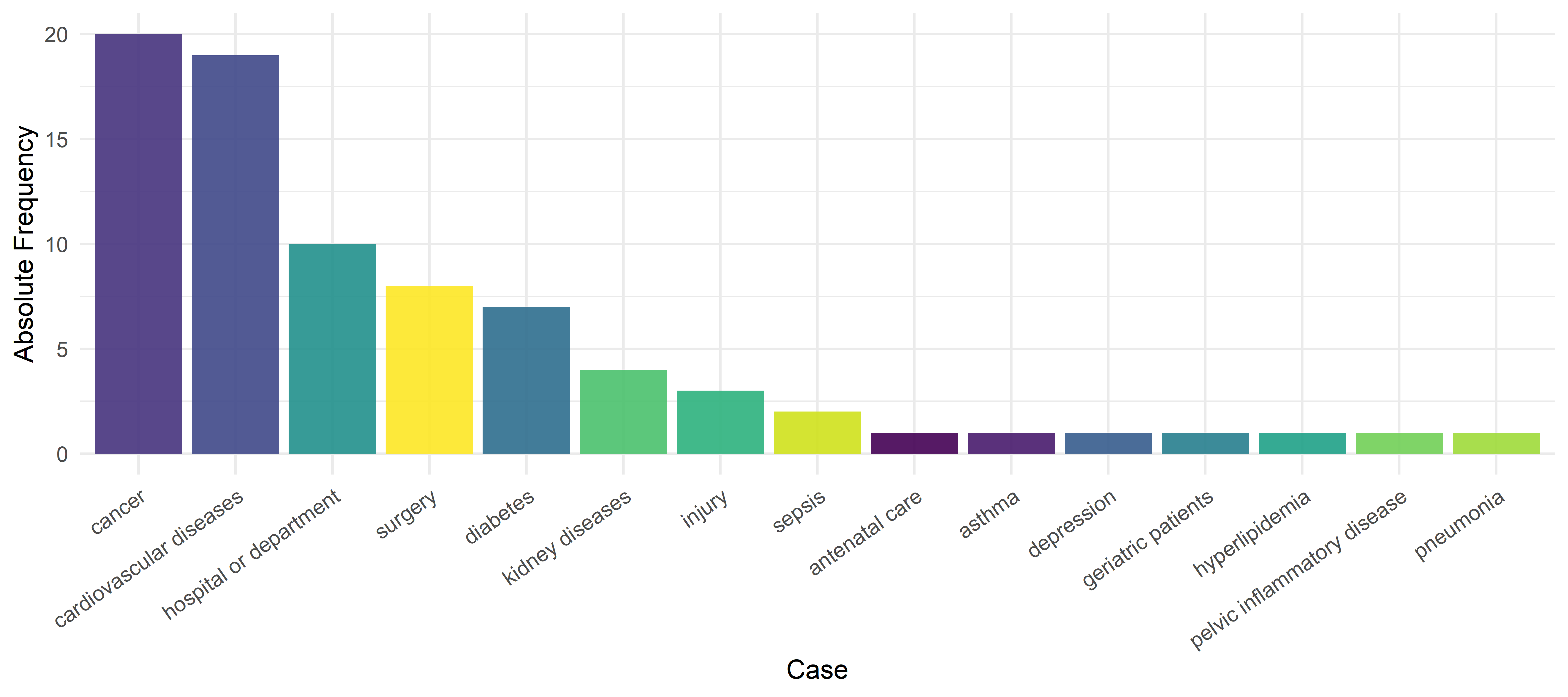}
    \caption{Frequency of illness or area of application}
    \label{fig:case}
\end{figure}

The reviewed articles have applied the automatic discovery of patient pathways to varied medical fields (Figure~\ref{fig:case}). The most frequent ones are cancer and cardiovascular diseases. 

Among cancer studies, breast cancer was one of the most investigated topics. \citet{Defossez2014} found three main pathways while studying breast cancer patients from a French region. These three patterns, which accounted for 95\% of the pathways, were: (i) surgery followed by radiotherapy; (ii) surgery followed by chemotherapy and radiotherapy; (iii) neoadjuvant therapy before surgery, followed by radiotherapy. The authors considered that all of them were in line with the recommendations. 
On the other hand, \citet{Zaballa2020} followed breast cancer patients from Spain using a clustering method. After assessing the central pathway of each cluster, they detected five patterns: (i) surgery followed by chemotherapy and radiotherapy; (ii) surgery followed by radiotherapy and then hormonal therapy; (iii) surgery followed by hospitalisation only; (iv) surgery followed by radiotherapy; and (v) neoadjuvant therapy before surgery, followed by hospitalisation and radiotherapy. 
It is noticeable that groups (iv), (i) and (v) are similar to the three patterns discovered by \citet{Defossez2014}, except for the presence of hospitalisations in pathway (v), which do not belong to the guidelines \cite{Zaballa2020}. 
Indeed, \citet{Dagliati2017} found that breast cancer pathways with one or more hospitalisations in the oncology ward tended to have higher values of cancer biomarkers (CEA and CA15-3), possibly indicating worsening patients. \citet{Chiudinelli2020} used an extension of their mining method to study breast cancer patients from northern Italy. The authors were able to identify three groups of pathways: (i) breast surgery possibly followed by plastic reconstruction; (ii) cancer recurrence leading to multiple surgeries; and (iii) neoadjuvant therapy before surgery. Instead of comparing biomarker values, they analysed the onset of metastases, local recurrences, and overall survival and found differences among the groups.

\citet{Dahlin2019_b} were also interested in breast cancer patients, but they assessed pathway costs and variability in four Swedish hospital groups. They concluded that the hospital that had more focus on developing and maintaining a theoretical clinical pathway had fewer pathway variants and also reduced time between diagnosis and surgery. \citet{Marazza2020} aimed at comparing pathways as well, but they focused on developing a method to compare pathway models. In their case study, they compared sub-populations with breast cancer and found that patients with a low likelihood of cancer (BIRADS 1-2) and screening patients had very similar pathways. In contrast, BIRADS 1-2 pathways presented higher difference scores when compared to patients with a high probability of malignancy (BIRADS 3-6).

Studies on other types of cancers used the strategy of comparing pathways too. \citet{Villamil2017} had the objective of assessing the quality of the Colombian health system, using gastric cancer as their case study. They chose two health service providers---an efficient and an inefficient one, according to the conducted Data Envelopment Analysis---and looked for differences in their patient pathways. The authors pointed out that it is a complex case, but it appears that the pathways from the less efficient unit are prone to have a second surgery, which could indicate flaws in the diagnostic phase. 
On the other hand, \citet{Kurniati2019a} discovered cancer patient pathways during their evaluation of the data quality of the Medical Information Mart for Intensive Care III database. They were able to compare the pathways before and after a change of the electronic health record system. The work of \citet{Bettencourt-Silva2015} is another example that evaluated data quality. The authors compared Prostate-Specific Antigen (PSA) trends with the patient pathways (containing treatment information). They investigated the moment and the frequency of the biomarker readings and whether the events of the patient pathway justified biomarker changes.

\citet{Antonelli2012} used sequence mining to discover the diagnostic pathways of colon cancer and found that, frequently, they did not follow the protocols. Among the identified pathway variants, they highlighted one in which the cancer diagnostic apparently occurs in the emergency room, which is undesirable because it might reflect an already critical state of the patient.   
\citet{Baker2017} also found multiple variants, a few of them following the theoretical clinical pathways for the treatment of breast cancer and colorectal cancer. The authors used a Markov Model to represent the pathways and used real data to learn the probabilities of moving from one state (pathway step) to another. Moreover, they investigated  the probabilities of some events of interest, such as telephone contact, the occurrence of neutropenia, and death. 
When studying pathways in the gynecologic oncology department, \citet{Caron2014} detected unexpected behaviour as well. However, they argue that it is not necessarily undesirable. For instance, although the list of cancers treated with paclitaxel did not include uterus, cervix and endometrium cancer, specialists of the studied hospital used it in those cases, which could mean they found a new use for the drug.

\citet{Stefanini2020a} assessed lung cancer patient pathways in a hospital to estimate the necessary resources to support their treatment. The authors found that the behaviour was somewhat homogeneous, usually consisting of some exams conducted before and after the lung surgery. \citet{Egho2015} analysed lung cancer patients as well, but they used multiple-perspective pathways, which included procedures, diagnoses and healthcare units. They clustered the patients into four groups and extracted representative patterns for each one. The first and the second groups involved patients who underwent a pneumonectomy but in two different French departments; the third group had multiple chemotherapy sessions in the same department as the first group; the fourth cluster had repeated chemotherapy sessions too but conducted in other regions. In previous work \cite{Egho2014}, the authors had also used lung cancer patient data to mine pathways but focused mainly on the proposed techniques.  Similarly, \citet{Aspland2021} used lung cancer patient pathways to evaluate the distance metric proposed by them. 

\citet{Huang2012} applied the pathway mining method they developed to four different types of cancer and two cardiovascular diseases. Their method consisted of mining frequent sequences and relevant time patterns. For bronchial cancer, they chose three activities and compared the time between them in the discovered model with the Ministry of Health recommendations and the hospital clinical pathway. They found that the time interval between surgery and discharge is longer in the mined pathway than in both guidelines. They argued the entities could use this information to improve the protocols. In further work \cite{Huang2013} the authors used a method based on representing the patient pathway as a set of time intervals with frequent patterns of treatments and applied it to four diseases, including bronchial cancer. Once again, they found that the time interval between surgery and discharge observed in the data differed from the ones specified in the guidelines.

The conformance of patient pathways to the guidelines was the objective of other cancer case studies too. \citet{Rinner2018} assessed the conformity of the follow-up process of melanoma patients and investigated if there was a correlation between it and the survival probabilities. They found that, for patients whose pathway was longer than two years, the survival probability was slightly higher when the pathways were compliant, but when they considered pathways that lasted less than two years, the opposite happened. They argued this might happen because severe patients have a shorter life expectancy and tend to follow the guidelines more carefully. 
\citet{Senderovich2016} evaluated the conformity of cancer patient pathways as well, but from an organisational perspective. They examined whether the processes in an oncology clinic were following the scheduled treatment and identified a bottleneck in the pharmacy station. 

The second medical field with more case studies encompassed cardiovascular diseases, such as stroke, heart failure and angina.
\citet{Gonzalez-Garcia2020} considered the pathways of patients who suffered a stroke. Because of the urgency of its care, their analyses included the assessment of the time between the onset of the stroke symptoms and the fibrinolysis treatment. The nine studied hospitals respected the maximum of 4.5 hours required by the guidelines. \citet{Sato2020} evaluated stroke patient pathways as well and found that some patients visited an Urgent Care Centre instead of going directly to a hospital, which increased the time between the onset of symptoms and hospital admission. 
\citet{Chen2018} studied the specific case of ischaemic stroke. They clustered the patients and, for each group, investigated the efficiency, costs and length of stay of each variant. They elected the optimal treatment process based on these criteria.
\citet{Leonardi2018} proposed the use of an ontology to increase the readability and facilitate the comparison of patient pathway models. In the case study, they compared stroke treatment in different healthcare units. They found that less equipped stroke units did not present some procedures, such as recanalization therapy and intra-cranial vessel inspection.  
\citet{Partington2015} have also compared patient pathways in different healthcare units. Their case study encompassed chest pain patients assisted in four hospitals. They found differences in the time between presentation and admission, but the proportion of patients considered urgent during the triage was also different across the units. 

\citet{Huang2015} used a method based on probabilistic topic models to find patients with similar patterns of care for unstable angina. They obtained three groups and found a correlation between unstable angina and hypertension in all of them. In another study \cite{Huang2015a}, the authors modified the method to find anomalies in the unstable angina pathways. They found that most of the local anomalies were delays in examinations or laboratory tests. In further work \cite{Huang2016}, the authors have included comorbidity information during the generation of the latent patterns and built a prediction model to estimate the provided treatment according to the set of comorbidities. 

\citet{Lin2001} have also conducted predictions based on comorbidities. They compared pathway models of stroke patients with different comorbidities and found that among their 113 patients, there were around 100 unique diagnostic codes. They used association analysis to predict the clinical path based on the set of diseases. 
\citet{Najjar2018} have also identified a variety of comorbidities in patients with cardiovascular diseases. The authors studied the case of heart failure, and they kept most of the records of these patients to find clusters of similar pathways, including the influence of comorbidities. They found, for instance, clusters of patients with respiratory problems, anaemia, renal failure and anxiety.

\citet{Prodel2018} followed the pathways of patients who had the implantation of a cardiac resynchronisation therapy defibrillator. The discovered pathway model agreed with the specialists' expectations. Moreover, they found that there was a 49.7\% of risk that the patients would have another heart failure on an average of eight months after the implantation.
\citet{Huang2018} grouped procedures and medications into topics, from which they built the patient pathways. They assessed patients with cardiovascular diseases and found that the two most frequent first topics contained physical examinations and lab tests, but one of them was more frequently used to treat urgent cases.
\citet{Hur2020} developed a system to explore patient pathways and mined twelve patterns related to unplanned cardiac surgeries. 

\citet{Zhang2018} mined pathways of three groups of patients: those with hypertension, type 2 diabetes and depression. For the three diseases, the variability of pathways may indicate that the treatment provided was not following the guidelines. Particularly, metformin (for diabetic patients) and amlodipine (for hypertension) did not appear as first-line therapy as much as expected. \citet{Xu2017} conducted two case studies, namely intracerebral haemorrhage, and inguinal hernia. The first group contains pathways with admission, followed by treatment and re-examination, while the latter involves examinations, pre-surgical activities and, eventually, the surgery for patients fit for it. They found that, on average, intracerebral haemorrhage patients required more nursing care than patients with inguinal hernia. \citet{Wang2017} conducted case studies on unstable angina pectoris, vertebrobasilar insufficiency and type 2 diabetes, but their discussion focused on the used methods. 

The third field with most case study applications does not refer to a specific disease but alludes to a hospital or department.

\citet{Rismanchian2017} used the mined patient pathways to optimise the layout of the emergency department of a South Korean hospital. They highlighted the importance of choosing the appropriate goal for the optimisation problem with an example from their findings: when the travel distances of non-critical and critical patients decreased by more than 44\% each, the design preference declined by 75\%. The authors proposed the use of a multi-objective problem to balance the multiple goals.\citet{Arnolds2018} also conducted a case study that involved layout optimising in a German hospital.

\citet{Rebuge2012} investigated patient pathways through an emergency department. They grouped the pathways and identified clusters of frequent and infrequent behaviour. Among the infrequent behaviour, they found, for instance, undesirable pathways in which an exam request appears after the scheduling and the exam itself. 
\citet{Stefanini2018} evaluated the patient pathways in an emergency department of a hospital as well. Among their analyses, they compared the pathways in summer and winter, given that the patient flow is usually higher in summer. Indeed, they found an increment in the waiting times between activities during summer, which affects the quality perceived by the patient. \citet{Benevento2019} aimed at predicting waiting times in an emergency department and found that the use of pathway information improves the accuracy of the prediction. 
Two additional articles used the emergency department as their case study \cite{Duma2020,Mertens2020}. 

\citet{Yoo2016} used process mining algorithms to discover patient pathways before  and after installing  a new building in a hospital. They evaluated performance indicators and found that the consultation wait time decreased with the new building, but the time spent by outpatients on administrative activities, including receipt of prescriptions, increased. In order to improve that, they recommended evaluating the possibility of reallocating the outpatient pharmacy. \citet{Kim2013} mined patient pathways to evaluate organisational processes in a hospital. The ten most frequent pathways agreed with the expectation of specialists. \citet{Lu2016} have also considered organisational patient pathways, but as a case study to validate the process discovery algorithm proposed by them. 

Surgery was the fourth medical field with more case studies. Two studies investigated appendectomy pathways. \cite{Yoo2015,Kempa-Liehr2020}. \citet{Yoo2015} examined its compliance and proposed enhancements to the clinical pathway, by removing recommended actions that were infrequent in the data and adding actions that were not present in the guidelines but were frequent in the observed pathways. On the other hand,  \citet{Kempa-Liehr2020} were interested in predicting the time between surgery and discharge based on the age of the patients, the surgery duration and the patient pathway variants. They found that the time tends to be longer for older patients in most of the variants. Nonetheless, they did not observe this behaviour for the two most frequent variants.

\citet{Cho2020} conducted two surgery-related case studies. The first one involved patients who underwent  total laparoscopic hysterectomy and the other comprised patients treating rotator cuff tears. They developed a method based on optimisation to discover the patient pathways to use frequent behaviour to propose or enhance clinical pathways. While analysing the results, they highlighted the importance of specialists' knowledge to check the mined models, because not everything frequent in the patient pathways should be made a rule in the guidelines. For example, although analgesic and digestive system medication were frequent, they are prescribed after surgery according to the necessity of the patients and should be not be indiscriminately given to every patient.

\citet{Erdogan2018b} analysed surgeries in a hospital from an organisational perspective. They detected that 65\% of the patients followed the same pathway and identified some bottleneck problems. \citet{Fei2013} also considered multiple types of surgeries, but their goal was to find groups of patients with similar speciality pathways. They found that some clusters were multidisciplinary, while others were related to specific surgeries. 
\citet{DeOliveira2020c} and \citet{Tamburis2020} also used surgeries as their case studies, but their discussion focused on the used methods.

Several case studies investigated diabetic patient pathways.
\citet{Conca2018} investigated the speciality pathways of diabetic patients. They found that more than 50\% of the patients did not receive support from the three recommended disciplines during one year (physician, nurse, dietitian). They also compared glycated haemoglobin trends among pathway groups and found statistically significant differences between them. \citet{Dagliati2018} also used glycated haemoglobin to compare diabetic patient pathways but using an intervention perspective. Similarly, different groups of patient pathways had different values of glycated haemoglobin (measured between the diagnosis and the first visit).

\citet{DeOliveira2020a} compared diabetic patient pathways from a diagnostic perspective based on the occurrence of four common complications, namely  stroke, infarction, amputation and terminal chronic kidney disease. They found that the time until the complication and the total time of the pathways differed among the groups. \citet{Khan2018} used a diagnostic perspective as well, but they had the objective of finding comorbidities more common in diabetic patients than in the general population. They found that cardiac arrhythmia, long term use of insulin, liver disease, cataract and valvular disease were frequently associated with the diabetic cohort. 

Among the analyses conducted by \citet{Lismont2016} on diabetic patients, they evaluated the education activities in the patient pathways. They found that more than 50\% followed a maximum of two educational activities per year, while other patients had more activities but without a long time interval between them.   

The remaining medical fields appeared in less than five case studies each and involved kidney diseases, injury, sepsis, antenatal care, asthma, depression, geriatric patients, hyperlipidemia, pelvic inflammatory disease and pneumonia. 

Concerning kidney diseases, two studies involved chronic kidney disease (CKD) \cite{Zhang2015a,ZhangPadman2015} and the other two comprised acute kidney injury (AKI) \cite{Zhang2015,Sawhney2021a}. 
\citet{ZhangPadman2015} built CKD patient pathways considering the following perspectives: visit type, diagnosis, medication and procedure. They clustered patients according to their pathways and found differences in medications and treatment intervals, even for patients with the same conditions. On the other hand, \citet{Zhang2015a} explored pathways aiming at predicting the chosen treatment and new events. 
While assessing AKI patient pathways, \citet{Zhang2015} reported that most of the pathways were aligned with what they expected, but some variants brought new information that should be validated in further work, such as an apparent improvement in patients who underwent educational sessions. On the other hand,  \citet{Sawhney2021a} identified that a significant number of patients who returned to the hospital after an AKI had no monitoring record in the previous two weeks. Thus, the authors stressed the necessity of improving readmission prevention.

Three studies investigated the pathways of injured patients. \citet{Andrews2020} examined trauma patients who had a road traffic accident. They considered three sub-cohorts of patients according to their level of interaction with the hospital and analysed the pathway models using a process mining algorithm. They were challenged by spaghetti models while assessing the cohort that involved hospitalisation and found that manually editing the pathway model with specialist guidance improved the results. 
\citet{Durojaiye2018} assessed trauma patient pathways from the perspective of the visited departments. They selected only patients classified with the two highest trauma activation levels. The authors were able to find that some pathways in the second most critical level followed pathways significantly different from the other patients in the group but quite similar to the pathways of patients from the most critical group. The authors reasoned that the triage team could have underestimated the health state of these patients and recommended investigating it. 
\citet{Mertens2018} studied patients with arm-related fractures but their work focused on the proposed method.

\citet{DeOliveira2020b} studied sepsis patient pathways from the perspective of diagnoses. They mined the pathways with an optimisation strategy and found many variants aligned with the specialists' expectations. They discovered some other patterns that might be relevant and should be investigated, such as the occurrence of recurrent urinary tract infections prior to the sepsis diagnosis. \citet{Prokofyeva2020a} also considered sepsis patients but built their pathways from a perspective of interventions. Their work focused on the proposed methods.

\citet{LeMeur2015} assessed the quality of antenatal care by extracting patient pathways from data as sequences and clustering them into three groups. The perspective used was a relative measure of the number of antenatal visits the patient had in each pregnancy trimester. They found that women from the cluster who had fewer antenatal visits were more likely to live in cities with higher unemployment rates and municipalities with a higher proportion of blue-collar workers when compared to the patients from the other groups.

\citet{Basole2015} developed an interactive tool to visualise patient pathways and used paediatric asthma patients arriving at the emergency department as their case study. They used information on the procedure, medication, administrative activities and health state.

\citet{Garg2009} studied the pathways of geriatric patients using the phase of treatment as the pathway perspective. They built a Markov Model and learnt its parameters. Among their findings, they detected that pathways with readmissions do not have a high occurrence probability and that the pathway variant related to the highest total expenditure was an individually inexpensive one that had a high probability of being followed. 

\citet{Perer2015} assessed the intervention pathways of patients with hyperlipidemia. They discovered frequent patterns using a sequence mining algorithm. The specialist who analysed the results was interested in patients who had either diabetes or hypertension as comorbidities. He identified that many pathways which resulted in undesirable LDL cholesterol levels contained drugs that could have contributed to it, such as specific antibiotic groups, glucocorticoids or fluoroquinolones.

\citet{Yang2006} considered the pathways of patients with pelvic inflammatory disease to evaluate a new method to detect healthcare fraud and abuse. 

\citet{Arias2020} conducted two case studies---pneumonia and acute myocardial infarction. They found that the patient pathways of both diseases usually start in the emergency room, and all patients with pneumonia required hospitalisation. They used a process mining software to analyse different variants and stratify the patients according to their age.

\subsubsection{Medical perspectives}

\begin{figure}[!ht]
    \centering
    \includegraphics[width=0.9\textwidth]{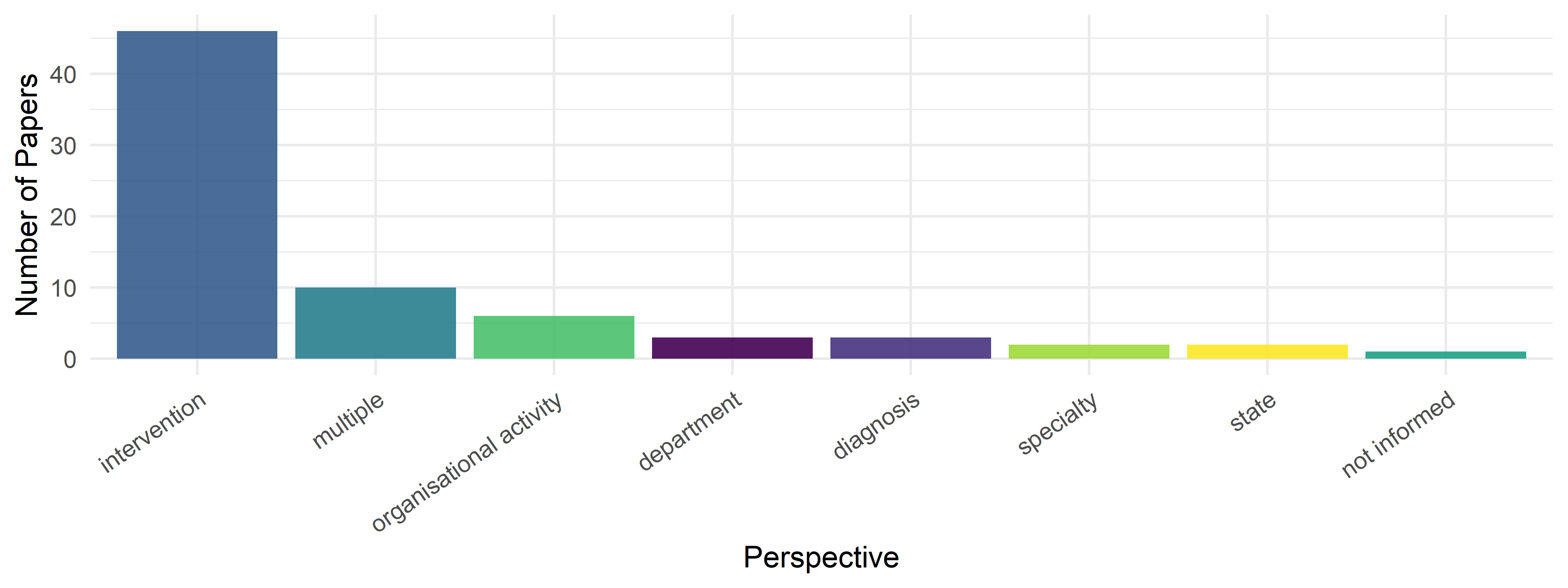}
   \caption{Frequency of patient pathways perspectives}
    \label{fig:perspective}
\end{figure}

Although patient pathways depict the trajectories followed by patients, the events that form the pathways can contain different types of information. As an example, one may consider the pathways as the historic of medical procedures a patient underwent, while other could take the sequence of consulted medical specialists. We identified seven different perspectives in the reviewed studies, as Figure~\ref{fig:perspective} depicts.

The most common perspective labelled `intervention' was used in more than 50\% of the papers. It encompasses medical procedures, drug administration and other related activities (e.g.\ admission and discharge). It is commonly used for assessing clinical pathways \cite{Cho2020,Gonzalez-Garcia2020}.

Six studies focused on organisational activities, such as scheduling, registration and payment \cite{Arias2020,Erdogan2018b,Kim2013,Lu2016,Rebuge2012,Yoo2016}. Thus, the health service area has benefited from patient pathway studies as well. The perspective `department' can reveal how patients move within a hospital \cite{Rismanchian2017}, or the departments can reflect the provided treatment \cite{Durojaiye2018}. 

Some authors built diagnosis pathways that represent the diseases the patients contracted or developed over time \cite{Prodel2018,DeOliveira2020b,Khan2018}, while others focused o the specialities the patients referred to \cite{Conca2018}. 

The treatment phase of elderly patients \cite{Garg2009} and an indicator of the number of visits a pregnant patient underwent in relation to the whole cohort \cite{LeMeur2015}  were labelled as `state'. 

The second largest category is actually a group of studies which considered multi-perspective pathways, as shown in Table~\ref{tab:multi_persp}.

\begin{table}[!ht]
    \centering
    \caption{Multi-perspective articles}
    \begin{tabular}{|c|c|}
    \hline
        \textbf{References} & \textbf{Perspectives} \\ \hhline{|=|=|}
        \cite{Dagliati2018,DeOliveira2020a,DeOliveira2020c} &  intervention and diagnosis \\ \hline
        \cite{Egho2014,Egho2015} &  intervention, diagnosis and unit or department \\ \hline
        \cite{Najjar2018} & intervention, diagnosis, unit or department and specialty\\ \hline
        \cite{Zhang2015,Zhang2015a,ZhangPadman2015} &  intervention, diagnosis and visit\\ \hline
        \cite{Fei2013} & unit or department and specialty\\ \hline
    \end{tabular}
    \label{tab:multi_persp}
\end{table}

\subsubsection{Employment of Patient Pathways}

The diversity of pathway models and mining techniques is also reflected on the variety of applications of patient pathways. One of the most common goals of the reviewed studies was to identify the actually-adopted clinical pathways. This could be done to verify whether the recommended guidelines were being followed \cite{Gonzalez-Garcia2020,Antonelli2012,Rinner2018}, to propose new guidelines \cite{Zhang2015,Chen2018}, or to compare medical practices in different healthcare units \cite{Partington2015,Leonardi2018}. Moreover, groups of patients can be identified and/or compared using their pathways \cite{Chiudinelli2020,Conca2018,Marazza2020}.

Patient pathways have also been used to evaluate organisational healthcare processes \cite{Erdogan2018b,Rebuge2012,Stefanini2018}, estimate the resources required by a group of patients \cite{Stefanini2020a}, and feed simulation models \cite{Tamburis2020}. The mobility of patients can be used to optimise the layout of healthcare units \cite{Arnolds2018,Rismanchian2017}. 

Some authors have mined patient pathways to evaluate the quality of a healthcare system \cite{Villamil2017,Sato2020} and the experience of patients with it \cite{Arias2020}. Other authors focused on the detection of fraud \cite{Yang2006} and the prediction of new events \cite{Zhang2015a} and time to discharge \cite{Kempa-Liehr2020}.

\section{Summary and Outlook} 
\label{sec:disc}

The first attempts to mine patient pathways have mostly used sequence mining or other discovery methods based on the Apriori algorithm. The most common patient pathway models included sequences, graphs and sets of time intervals. Although the used methods could discover frequent patterns effectively, relevant infrequent behaviour may have been overlooked. The capability of the pathway models to represent relations between activities was usually limited to direct successions, but most of the models used allowed repeated activity labels, which improved the readability. The notion of time was usually restricted to the order of the events. 

With the development of process mining as a research area and its application in healthcare, the expressiveness of the models increased, 
brought by the identification of  joins and splits of the processes and long-term dependencies, for instance. Additionally, most process mining tools provide a summary of the time elapsed between activities. If on the one hand, the models could be more expressive, on the other, the challenge of dealing with the high variability of pathways and activities became more evident. Traditional process mining algorithms focus on discovering structured process models, and hence, they usually confront spaghetti processes when dealing with healthcare data. Moreover, several of them do not support repeated activity labels, which can hinder the interpretation of the model. Alternatively, auxiliary methods, such as clustering events or patients, and filtering events or pathways have been adopted. 

Patient clustering has been used not only as an auxiliary method, but also as the main strategy for assessing patient pathways, either taking a central pathway as representative of the cluster or evaluating the temporal distribution of activities within each cluster. Although it can deal well with data variability, it does not generate an explicit pathway model. 

Other used approaches focused on specific cases, such as updating Markov models or simply extracting pathways as they were on data.

Meanwhile, within the process mining domain, other discovery algorithms have been proposed, focusing on less structured processes, such as the healthcare ones. An example is the Fuzzy Miner algorithm which uses a graph as a process model and simplifies it by identifying relevant nodes and edges. Thus, the pathway variability can decrease without necessarily removing significant infrequent behaviour. Many tools have implemented this simplification approach, including commercial ones, and it has become popular for mining patient pathways. Nevertheless, the use of a graph that does not allow repeated activity labels and only shows directly-follows relationships, combined with implementations of discovering algorithms that simply filter nodes and edges, can lead to unclear and even misleading models, as \citet{VanDerAalst2019} has recently discussed. 
Moreover, although the idea proposed in the Fuzzy Miner algorithm was to select relevant nodes and edges, it seems that most authors who use this approach limit it to identifying frequent activities and transitions.

More recently, optimisation problems have been proposed to obtain patient pathway models. This strategy is somewhat similar to the simplification one because it also looks for key nodes and edges to be included in the final model, but it is guided by an objective function, such as the value of a metric. Depending on the mathematical model used, these algorithms can face the same issues as the simplification ones, but alternative representations have been proposed, such as the Time grid process model of \citet{DeOliveira2020a}, which distinguishes activities occurring in different moments of a pathway. The authors who used this approach focused mainly on maximising the replayability measure of the model. In other words, their algorithms chose activities and transitions such that the pathway model could reproduce as much as possible the observed behaviour, which might favour frequent occurring patterns. 

The review of the modelling aspects of the papers suggests that the development of a mining method that deals with data variability by selecting relevant activities and transitions based on their importance for the pathway model and the time and context influence would be advantageous. It should also provide a pathway model that is easy to read and interpret.
This is a challenging and still open topic in the literature.



\bibliography{Modelling_and_Mining_of_Patient_Pathways_A_Scoping_Review}


\end{document}